\documentclass[12pt]{article}
\usepackage{amssymb}
\usepackage{epsfig,caption,graphicx,eepic,epic}
\usepackage{pstricks}
\usepackage{amsmath}
\usepackage{multido}
\usepackage{array}
 
\begin{document}
\def\be{\begin{equation}}
\def\ee{\end{equation}}
\def\ba{\begin{eqnarray}} 
\def\ea{\end{eqnarray}}
\def\nn{\nonumber}

\newcommand{\bbf}{\mathbf}
\newcommand{\rrm}{\mathrm}

\title{Excitation of time-dependent quantum systems: an application
of time-energy uncertainty relations\\} 

\author{Tarek Khalil$^{a}$
\footnote{E-mail address: tkhalil@ul.edu.lb}\\ 
and\\
Jean Richert$^{b}$
\footnote{E-mail address: j.mc.richert@gmail.com}\\ 
$^{a}$ Department of Physics, Faculty of Sciences(V),\\
Lebanese University, Nabatieh,
Lebanon\\
$^{b}$ Institut de Physique, Universit\'e de Strasbourg,\\
3, rue de l'Universit\'e, 67084 Strasbourg Cedex,\\ 
France} 

\maketitle 
\begin{abstract}
The conditions under which time-energy uncertainty relations derived by Deffner and Lutz~\cite{deflu} 
for time-dependent quantum systems minimize the time necessary to excite such systems from their ground 
state to excited states are examined. The generalized Margolus-Levitin and Mandelstam-Tamm inequalities are 
worked out for specific fermionic and bosonic systems.
\end{abstract} 
\maketitle

\section{Introduction}

Time is an essential concept for the realization of fast electronic devices. As an example it may be useful 
to control and minimize the time needed by a two-level quantum system to jump from its ground state to
an excited state. This time is constrained by Heisenberg's time-energy uncertainty relation and bounded by a lower amount known as a quantum speed limit(QSL) time.\\

The optimization of the time duration of quantum jumps for stationary systems has been the object of numerous theoretical studies
starting with Mandelstam and Tamm~\cite{tm} and pursued up to present time in order to attain the sharper lower bound~\cite{lv,ju,ml,lt,jk}.\\



A vast litterature has been devoted to various aspects of the problem of time dependent systems~\cite{aa,pp,and,deflu}.
 Anandan and Aharonov have used alternative geometric derivations to obtain expressions for the Fubini-Study metric where the shortest possible distance between orthogonal states, which is along a geodesic, leads to get implicit bounds for the time of evolution of a quantum system~\cite{aa,pp}. Other authors have used differential geometric methods to get sharper uncertainty relations for mixed states that can be optimized in the case for fully distinguishable states. Moreover they characterize the Hamiltonians that optimize the evolution time for finite-level quantum systems~\cite{and}. Recently for arbitray quantum unitary processes Deffner and Lutz ~\cite{deflu} extended the Mandelstam-Tamm(MT) and the Margolus-Levitin (ML) inequalities to time dependent systems which are either intrinsically time-dependent or driven by an external time-dependent perturbation. To this end, they derive the upper bounds for the Bures angle(Bures length) or, rather, for Fisher information~\cite{jk}. At this stage we optimize the time in the extended ML formulation with respect to a maximum value. Besides, this principle of optimization, that fixes the minimum time, is also applicable for all the estimates presenting an extremum. Another groups have explored the MT bound, which is geometric in nature, to attain the minimum time of evolution in the context of the time-optimal control(time-OC) problem~\cite{pgi,cnv,hgf,def}. More specifically, this method seems useful to minimize the decoherence for a system~\cite{bas}. Further works have been developed for both unitary and non-unitary processes~\cite{deflu1,tad,cap}.\\

In practice it is of interest to test the time needed by a quantum system to be driven from its ground state 
into excited states. In the present work the inequalities derived in~\cite{deflu} have been used and worked out in 
order to determine the time needed in order to generate quantum transitions.
In the first part of the present work we examine the general conditions under which the ML inequality is 
optimized i.e. comes closest to an equality for specific values of the time interval over which the system 
evolves. In a second step the ML and MT relations are applied to fermionic and bosonic systems.\\

In section 2 the inequalities are explicitly formulated and the conditions under which the inequalities 
can be optimized in the ML case are presented in section 3. Section 4.1 is devoted to a quantitative study of the
MT and ML expressions by applying them to a fermionic $1d$ quantum chain. In section 4.2 the ML inequality is 
applied to a simple bosonic system coupled to an external time-dependent perturbation which can act in a weak or 
strong coupling regime.

\section{Time-energy inequalities}
 
The general structure of the time-energy inequality can be written as~\cite{deflu} 

\ba
\tau \geq R(\tau)
\label{eq1}   
\ea
 where $R(\tau)=\frac{\hbar C_{0\tau} }{\Delta E(\tau)}$, $C_{0\tau}=\arccos \Omega_{0\tau}$, $\Omega_{0\tau}=|\langle \Psi(0) |\Psi(\tau) \rangle|$ is the overlap between the wave function at time
$t=0$ and $t=\tau$, and $\Delta E(\tau)$ characterizes the energy difference 
acquired by the system between the initial and final time. In the following $\hbar=1$.\\


In the  Mandelstam-Tamm formulation the energy denominator $\Delta E(\tau)$ of Equation (1) is given in terms of 
the variance of the energy

\ba
\Delta E(\tau)=1/\tau \int_{0}^{\tau}dt\frac{[<H(t)^{2}>-<H(t)>^{2}]^{1/2}}{\langle \Psi(t)|\Psi(t)\rangle}
\label {eq2}
\ea

with 

\ba
<H(t)^{2}>=\langle \Psi(t)|H(t)^{2}|\Psi(t)\rangle
\label {eq3}
\ea


In the Margolus-Levitin formulation $\Delta E(\tau)=<E(\tau)>- E(0)$ where $E(0)$ is the energy at $t=0$
and $<E(\tau)>= 1/\tau \int E(t) dt$  is the average energy of the system over a time interval $[0,\tau]$.

\section{Extrema in the Margolus-Levitin expression}

In the case of a time-independent system the realization of eq.(1) as a strict equality can be of great 
practical interest since it leads to the determination of the minimum time needed by the system starting 
from its ground state to excited states. This question has been successfully answered in ~\cite{lt,aa}.\\

The question may also be raised in the case where the hamiltonian dynamics are time-dependent when it is of
interest to find out the minimal time interval $[0,\tau]$ needed in order to realize the equality. It comes 
out that an analytic solution has not been found yet. A priori a less ambitious empirical answer might 
correspond to $C_{0\tau}=0$ and a minimum of the energy denominator $\Delta E(\tau)$.\\

A rigorous answer to the optimization of the inequality towards an equality consists in a determination of the 
maximum of $R(\tau)$ which brings the r.h.s. of the expression closest or equal to $\tau$.\\
 
The first derivative of this quantity with respect to $\tau$ leads to 

\ba
dR(\tau)/d\tau= \frac{\hbar[\Delta E(\tau)dC_{0\tau}/d\tau - C_{0\tau}
d\Delta E(\tau)/d\tau]}{\Delta^{2} E(\tau)}
\label {eq4} 
\ea

An extremum is reached if $dR(\tau)/d\tau=0$ which leads to

\ba
\frac{dC_{0\tau}/d\tau}{C_{0\tau}}=\frac{d\Delta E(\tau)/d\tau}{\Delta E(\tau)}
\label {eq5}
\ea

This extremum is a maximum if, for the corresponding value of $\tau$

\ba
\Delta E(\tau) d^{2}{C_{0\tau}}/d\tau^{2} -C_{0\tau} d^{2}\Delta E(\tau)
/d\tau^{2}<0
\label {eq6}   
\ea

A consequence of Eq.(5) can be observed if $C_{0\tau}$ is maximized. This corresponds to
$\Omega_{0\tau}=|\langle \Psi(0) |\Psi(\tau) \rangle|=0$, the vectors $|\Psi(\tau) \rangle$ and $|\Psi(0) \rangle$ 
are orthogonal to each other. Then $dC_{0\tau}/d\tau=0$ and induces $d\Delta E(\tau)/d\tau=0$ if 
$\Delta E(\tau) \neq 0$. But, in the ML formulation

\ba
d\Delta E(\tau)/d\tau=\frac{E(\tau)}{\tau}-\frac{<E(\tau)>}{\tau}=0
\label {eq7}   
\ea

which leads back to the expression of $\Delta E(\tau)$ if $E(0)=0$. Hence the stationarity of $\Delta E(\tau)$ 
is correlated with the orthogonality of the vectors $|\Psi(\tau) \rangle$ and $|\Psi(0) \rangle$. The stationary point 
$\tau$ can be an inflexion point or an extremum. Then $C_{0\tau}= \pi/2$ $mod(k\pi)$ and 

\ba
\tau \geq \frac{\hbar \pi/2 }{\Delta E(\tau)} 
\label {eq8} 
\ea
Whether or not this limit can be reached and $\geq$ replaced by a strict equality depends on 
the system.\\

\section{Models and applications}

\subsection{Fermionic 1d chain}

In a first step the time-energy inequality is applied to the time-dependent $1d$ chain with even periodic boundary 
conditions already introduced in ~\cite{bar,kat} 

\ba
H_{0} = J/2 (1+\gamma)\sum_{(i)} \sigma^{x}_{i}\sigma^{x}_{i+1} + J/2 (1-\gamma)\sum_{(i)} \sigma^{y}_{i}\sigma^{y}_{i+1}
- h_{0}\sum_{(i)}\sigma^{z}_{i}
\label{eq9}   
\ea 

where $\sigma^{x}_{i}$ is the $x$ component of the Pauli matrix and similarly for the $y$ and $z$ components.
The system is integrable and the wave function is given as a product of single particle wave functions with 
corresponding energies~\cite{jt}. The time dependence is generated by a local excitation of the last spin by an 
external magnetic field  

\ba
H^{(N)}_{1}(t)= h_{1}\exp(-t/\tau_{H}) S^{z}_{N}
\label {eq10}
\ea
with $S^{z}_{N}=\sigma^{z}_{N}/2$~\cite{jt}.\\

The wave function of the system is obtained perturbatively, up to second order in the interaction which works 
as a perturbation and leads to the expression of the overlap

\ba
\Omega_{0\tau}=|\langle\Psi(0)|(1+U^{(1)}(0,\tau)+U^{(2)}(0,\tau))|\Psi(0)\rangle|
\label {eq11}
\ea

where $|\Psi(0)\rangle$ is the wave funtion at $t=0$ and $U^{(1)}(0,\tau)$, $U^{(2)}(0,\tau)$ are the first and second 
order contribution to the evolution operator $U(0,\tau)=exp[-i(H_{0}\tau+\int_{0}^{\tau} dtH^{(N)}_{1}(t))]$. 
A justification for the neglect of higher order contributions used in the numerical application is given in the 
Appendix.\\

The energy of the system over a time interval $[0,\tau]$ reads

\ba
E(\tau)= 1/\tau \int_{0}^{\tau}\langle \Psi(t)|H(t)|\Psi(t)\rangle/\langle \Psi(t)|\Psi(t)\rangle-E(0)
\label {eq12}
\ea
where $|\Psi(t)\rangle$ is the perturbed wave function evaluated up to order $2$ and $E(0)$ the initial energy of the system.

\subsubsection{General considerations concerning the application of the model} 

The present external magnetic field on an unique spin state of the chain has been explicitly chosen to produce a weak 
effect on the chain in order to allow for a perturbative treatment. Under these conditions and for fixed $J=1$ it comes 
out that the wave function overlaps are not very sensitive to the strength of the magnetic fields , neither $h_{0}$ nor 
$h_{1}$ as long as these quantities stay in the range of unity. The same is true in the case of variations of the asymmetry 
parameter $\gamma$ which are fixed in the interval $[0,1]$. In all applications the length of the chain is $N=100$.

\subsubsection{Application of the Mandelstam-Tamm expression} 

Typical results concerning the r.h.s. of the MT expression are shown in Figure 1 as function of $\gamma$ with $\tau=100$, 
$h_{0}=1$, $h_{1}=1$. The results show that $R(\tau)$ keeps very small when compared with $\tau$. 


\subsubsection{Application of the Margolus-Levitin expression} 

Using the ML expression leads to the results shown in Table 1 for $R(\tau$). Here $\tau=100$, $h_{0}=1$, 
$h_{1}=1$ in lines (1-2), $h_{0}=1$, $h_{1}=2$ in line 3. $R(\tau)$ keeps again quite small when compared with $\tau$. 


\subsubsection{General comments} 
 
As a consequence of the perturbative nature of the time dependence of the total Hamiltonian $H(t)=H_{0}+H^{(N)}_{1}(t)$
the overlap between the wave functions at $t=0$ and $t=\tau$ staying in the numerator of $R(\tau)$ keeps relatively sizable, 
hence $C_{0\tau}$ never gets close to its maximum $\pi/2$ which would correspond to orthogonal states.\\

The energy denominators in both the MT and ML expressions are at least an order of magnitude larger that the numerator
so that $R(\tau)$ gets a small quantity in every investigated case. Since the external excitation is confined to a small
time interval the evolution of this quantity reaches quickly a quasi-stationary value after a very short number of time 
steps. The time-energy inequality is strictly verified.




\subsection{Bosonic model} 

As a second application consider a simple time-dependent bosonic system with $0$ and $1$ quantum excitation. The Hamiltonian 
is given by the expression

\ba
H=A(a^{+}a+1/2)+V^{*}(t)a^{+}+V(t)a
\label {eq13} 
\ea

where $V(t)=V_{0}e^{i\omega t}$ is an external perturbation. The wave function obeys the time-dependent Schroedinger 
equation and it is written as  

\ba
|\Psi(t)\rangle= \sum_{n} c_{n}(t)e^{-iE_{n}t}|n>
\label {eq14} 
\ea

Here one keeps two states, the ground state $|n>=|0>$ and an oscillator quantum excitation $a^{+}|0>$. Then the normalized 
amplitudes $c_{n}(t)$ can be determined analytically as solutions of a system of two coupled first order differential equations 

\ba
dc_{0}(t)/dt=-ie^{-i\omega t} V(t)c_{1}(t)/\hbar
\\ \nn
dc_{1}(t)/dt=-ie^{i\omega t} V^{*}(t) c_{0}(t)/\hbar
\label {eq15} 
\ea

The second order differential equation obtained for $c_{0}(t)$ shows three types of solutions 
depending on the sign of the determinant 

\begin{center} 
$\Delta= 4V^{2}_{0}/\hbar^{2}-(\omega-A)^{2}$.
\end{center}

The amplitudes $c_{0}(t)$ and  $c_{1}(t)$ show an oscillatory behaviour if $\Delta <0$ or equal $0$ 
and a product of an exponential and oscillatory behaviour when $\Delta>0$.

The knowledge of the time-dependent wave function 

\ba
|\Psi(t)\rangle= \bar c_{0}(t)e^{-iE_{0}t}|0>+\bar c_{1}(t)e^{-iE_{1}t}a^{+}|0>
\label {eq16}
\ea

where $E_{0}=A/2$, $E_{1}=3A/2$ and $\bar c_{0}(t)$, $\bar c_{1}(t)$ are the normalized amplitudes. 
The overlap $C_{0\tau}$ and energy $E(\tau)$ can be determined for any value of $\tau$
 
\ba
\langle \Psi(0)|\Psi(\tau)\rangle=(R_{0}(0,\tau)+iI_{0}(0,\tau))e^{-iE_{0}\tau}+(R_{1}(0,\tau)+iI_{1}(0,\tau))
e^{-iE_{1}\tau}
\label {eq17}
\ea
where $R_{0}(0,\tau)$ and $I_{0}(0,\tau)$ are the real and imaginary parts of $\bar c_{0}^{*}(0)\bar c_{0}(\tau)$
and similarly for $R_{1}(0,\tau)$ and $I_{1}(0,\tau)$. Then the expression of the energy reads

\ba
E(\tau)=1/\tau \int_{0}^{\tau}A(|\bar c_{1}(t)|^{2}+1/2)+2Re[V^{*}(t)\bar c_{1}^{*}(t)\bar c_{0}(t)
e^{i(E_{1}-E_{0)}t}dt]-E(0)
\label {eq18}
\ea

\subsubsection{General considerations concerning the model} 

Here the strength of the coupling in the time-dependent part of the Hamiltonian can be weak as well as strong:\\

Three different regimes can be generated. They correspond to $\Delta>0$, $\Delta=0$, $\Delta<0$.

\subsubsection{Analysis of the Margolus-Levitin inequality in these different regimes}

Here the initial conditions are fixed as $c_{0}(0)$=1, $dc_{0}$/dt=0.



\begin{itemize}

\item If $\Delta<0$ (weak coupling) one can come close to the maximum of $C_{0\tau}$ which leads also
      closest to a state which is not far from being orthogonal to the initial state $|\Psi(0)\rangle$. Calculations 
      indicate that $R(\tau)$ reaches a maximum when $C_{0\tau}$ gets to a maximum. This is in qualitative agreement
      with the considerations of section 3 about the fact that the denominator $\Delta E(\tau)$ of $R(\tau)$ can reach 
      an extremum when its numerator is maximum.

Some values of $R(\tau)$ are shown in Figure 2. Here $A=1$, $\omega=2$, $V_{0}=0.475$. The oscillations  
are due to the oscillating nature of the external perturbation.


\item If $\Delta \ge 0$ (intermediate and strong coupling) the situation is similar to the one observed in
      the fermionic chain system. The final state $\Psi(\tau)$ reaches a maximum of $\pi/4$, $C_{0\tau}$   
      and $R(\tau)$ reach very quickly a stationary value. This can be explained by the structure of 
      the wave function which stabilizes to a final value due to an exponential time component that multiplies
      an oscillatory term.\\

Some values of $R(\tau)$ are shown in Figure 3. Here $A=6$, $\omega=4$, $V_{0}=3$. 


\end{itemize}

In all cases the time-energy inequality is numerically verified.\\

\section{Conclusions}

The extension of the Margolus-Levitin and Mandelstam-Tamm time-energy inequality to time-dependent systems has been 
investigated on hand of two models, a fermionic $1d$ chain and a two-state bosonic system. The numerical
calculations are in agreement with the analytical expressions, a fixed time interval is always larger than the r.h.s.
of the inequality expressions.\\

In the case of the Margolus-Levitin formulation it has been shown that the minimum overlap between the wave function
at the initial time and the time evolved wave function reaches a minimum when the r.h.s.of the inequality is itself  
at an extremum. Numerical calculations confirm this point in the case where this extremum corresponds to an exact orthogonality 
or not. \\
 

\section{Appendix}


The perturbation induced by the external perturbation given in eq. (10) is essentially governed by the relaxation
time $\tau_{H}$ and the field strength $h_{1}$. To first order the strength of the perturbation is fixed by
$\tau_{H}h_{1}$. At second order the magnitude of the different contributions are governed by the strength factors 

\ba
S_{1} = h_{1}^{2} \tau_{H}^{2}
\\ \nn
S_{2} = h_{1}^2  (1/D^2)
\\ \nn
S_{3} = h_{1}^2/(\tau_{H} D)
\\ \nn
D \sim (1/\tau_{H}^2 + \Delta \epsilon^2)
\label{eq19}   
\ea 
where $\Delta \epsilon \sim \epsilon^{(i)}-\epsilon^{(j)}$, $(i,j)$ corresponding to single particle
ground or excited states. The single particle state energies are of the order of unity. In the numerical 
applications $h_{1}$ is chosen to be of the same order of magnitude and $\tau_{H}$ is two to three orders of magnitude 
smaller. As a consequence the second order contributions are two to three orders of magnitude smaller than the 
zeroth order ones. This justifies the cut-off of the expansion at order $2$.\\

{\bf\Large{Acknowledgments}}\\

The authors would like to thank Eric Lutz for helpful comments and advices.

\begin{figure}
\begin{center}
\setlength{\unitlength}{0.240900pt}
\ifx\plotpoint\undefined\newsavebox{\plotpoint}\fi
\sbox{\plotpoint}{\rule[-0.200pt]{0.400pt}{0.400pt}}%
\begin{picture}(1500,1350)(0,0)
\sbox{\plotpoint}{\rule[-0.200pt]{0.400pt}{0.400pt}}%
\put(160.0,1123.0){\rule[-0.200pt]{4.818pt}{0.400pt}}
\put(140,1123){\makebox(0,0)[r]{0.099}}
\put(1419.0,1123.0){\rule[-0.200pt]{4.818pt}{0.400pt}}
\put(160.0,869.0){\rule[-0.200pt]{4.818pt}{0.400pt}}
\put(140,869){\makebox(0,0)[r]{0.084}}
\put(1419.0,869.0){\rule[-0.200pt]{4.818pt}{0.400pt}}
\put(160.0,649.0){\rule[-0.200pt]{4.818pt}{0.400pt}}
\put(140,649){\makebox(0,0)[r]{0.071}}
\put(1419.0,649.0){\rule[-0.200pt]{4.818pt}{0.400pt}}
\put(160.0,445.0){\rule[-0.200pt]{4.818pt}{0.400pt}}
\put(140,445){\makebox(0,0)[r]{0.059}}
\put(1419.0,445.0){\rule[-0.200pt]{4.818pt}{0.400pt}}
\put(160.0,276.0){\rule[-0.200pt]{4.818pt}{0.400pt}}
\put(140,276){\makebox(0,0)[r]{0.049}}
\put(1419.0,276.0){\rule[-0.200pt]{4.818pt}{0.400pt}}
\put(1439.0,123.0){\rule[-0.200pt]{0.400pt}{4.818pt}}
\put(1439,82){\makebox(0,0){1}}
\put(1439.0,1290.0){\rule[-0.200pt]{0.400pt}{4.818pt}}
\put(1183.0,123.0){\rule[-0.200pt]{0.400pt}{4.818pt}}
\put(1183,82){\makebox(0,0){0.8}}
\put(1183.0,1290.0){\rule[-0.200pt]{0.400pt}{4.818pt}}
\put(800.0,123.0){\rule[-0.200pt]{0.400pt}{4.818pt}}
\put(800,82){\makebox(0,0){0.5}}
\put(800.0,1290.0){\rule[-0.200pt]{0.400pt}{4.818pt}}
\put(416.0,123.0){\rule[-0.200pt]{0.400pt}{4.818pt}}
\put(416,82){\makebox(0,0){0.2}}
\put(416.0,1290.0){\rule[-0.200pt]{0.400pt}{4.818pt}}
\put(160.0,123.0){\rule[-0.200pt]{0.400pt}{4.818pt}}
\put(160,82){\makebox(0,0){0}}
\put(160.0,1290.0){\rule[-0.200pt]{0.400pt}{4.818pt}}
\put(160.0,123.0){\rule[-0.200pt]{308.111pt}{0.400pt}}
\put(1439.0,123.0){\rule[-0.200pt]{0.400pt}{285.948pt}}
\put(160.0,1310.0){\rule[-0.200pt]{308.111pt}{0.400pt}}
\put(160.0,123.0){\rule[-0.200pt]{0.400pt}{285.948pt}}
\put(-1,716){\makebox(0,0){                       {\it R }              }}
\put(799,21){\makebox(0,0){$\gamma$}}
\put(1279,1270){\makebox(0,0)[r]{$\tau_{H}=0.001$}}
\put(416,445){\raisebox{-.8pt}{\makebox(0,0){$\Diamond$}}}
\put(800,411){\raisebox{-.8pt}{\makebox(0,0){$\Diamond$}}}
\put(1183,276){\raisebox{-.8pt}{\makebox(0,0){$\Diamond$}}}
\put(1349,1270){\raisebox{-.8pt}{\makebox(0,0){$\Diamond$}}}
\put(1279,1229){\makebox(0,0)[r]{$\tau_{H}=0.01$}}
\put(416,1123){\makebox(0,0){$+$}}
\put(800,869){\makebox(0,0){$+$}}
\put(1183,649){\makebox(0,0){$+$}}
\put(1349,1229){\makebox(0,0){$+$}}
\put(160.0,123.0){\rule[-0.200pt]{308.111pt}{0.400pt}}
\put(1439.0,123.0){\rule[-0.200pt]{0.400pt}{285.948pt}}
\put(160.0,1310.0){\rule[-0.200pt]{308.111pt}{0.400pt}}
\put(160.0,123.0){\rule[-0.200pt]{0.400pt}{285.948pt}}
\end{picture}
\end{center}
\caption{$R(\tau)$ as a function of $\gamma$ for $\tau$=100.}
\end{figure}
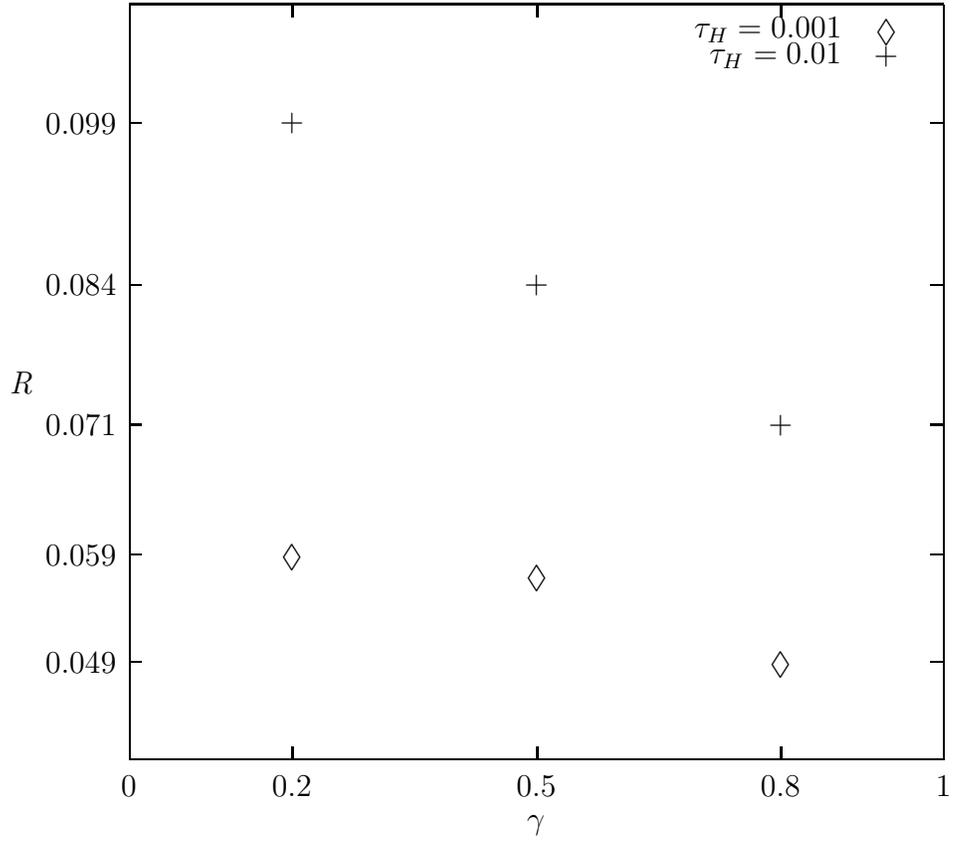

\begin{table}
\centering
\begin{tabular}{|l|c|r|}
\hline
  $\gamma$ & $\tau_{H}$ &  $R(\tau)$\\ 
\hline
  0.1 & 0.01 & 0.16 \\
\hline
  0.2 & 0.01 & 0.34 \\ 
\hline 
  0.5 & 0.01 & 0.04 \\
\hline
\end{tabular}
\caption{\label{Table 1}  $R(\tau)$ as a function of $\gamma$ and time $\tau_{H}$}
\end{table} 

\begin{figure}
\begin{center}
\setlength{\unitlength}{0.240900pt}
\ifx\plotpoint\undefined\newsavebox{\plotpoint}\fi
\sbox{\plotpoint}{\rule[-0.200pt]{0.400pt}{0.400pt}}%
\begin{picture}(1500,900)(0,0)
\sbox{\plotpoint}{\rule[-0.200pt]{0.400pt}{0.400pt}}%
\put(160.0,813.0){\rule[-0.200pt]{4.818pt}{0.400pt}}
\put(140,813){\makebox(0,0)[r]{1.17}}
\put(1419.0,813.0){\rule[-0.200pt]{4.818pt}{0.400pt}}
\put(160.0,707.0){\rule[-0.200pt]{4.818pt}{0.400pt}}
\put(140,707){\makebox(0,0)[r]{0.99}}
\put(1419.0,707.0){\rule[-0.200pt]{4.818pt}{0.400pt}}
\put(160.0,347.0){\rule[-0.200pt]{4.818pt}{0.400pt}}
\put(140,347){\makebox(0,0)[r]{0.38}}
\put(1419.0,347.0){\rule[-0.200pt]{4.818pt}{0.400pt}}
\put(160.0,253.0){\rule[-0.200pt]{4.818pt}{0.400pt}}
\put(140,253){\makebox(0,0)[r]{0.22}}
\put(1419.0,253.0){\rule[-0.200pt]{4.818pt}{0.400pt}}
\put(1237.0,123.0){\rule[-0.200pt]{0.400pt}{4.818pt}}
\put(1237,82){\makebox(0,0){200}}
\put(1237.0,840.0){\rule[-0.200pt]{0.400pt}{4.818pt}}
\put(900.0,123.0){\rule[-0.200pt]{0.400pt}{4.818pt}}
\put(900,82){\makebox(0,0){150}}
\put(900.0,840.0){\rule[-0.200pt]{0.400pt}{4.818pt}}
\put(564.0,123.0){\rule[-0.200pt]{0.400pt}{4.818pt}}
\put(564,82){\makebox(0,0){100}}
\put(564.0,840.0){\rule[-0.200pt]{0.400pt}{4.818pt}}
\put(227.0,123.0){\rule[-0.200pt]{0.400pt}{4.818pt}}
\put(227,82){\makebox(0,0){50}}
\put(227.0,840.0){\rule[-0.200pt]{0.400pt}{4.818pt}}
\put(160.0,123.0){\rule[-0.200pt]{308.111pt}{0.400pt}}
\put(1439.0,123.0){\rule[-0.200pt]{0.400pt}{177.543pt}}
\put(160.0,860.0){\rule[-0.200pt]{308.111pt}{0.400pt}}
\put(160.0,123.0){\rule[-0.200pt]{0.400pt}{177.543pt}}
\put(19,491){\makebox(0,0){                      {\it R }              }}
\put(799,21){\makebox(0,0){$\tau$}}
\put(248,813){\makebox(0,0)[l]{(1.44)}}
\put(584,253){\makebox(0,0)[l]{(0.28)}}
\put(921,707){\makebox(0,0)[l]{(1.22)}}
\put(1257,347){\makebox(0,0)[l]{(0.47)}}
\put(1279,820){\makebox(0,0)[r]{weak coupling}}
\put(227,813){\raisebox{-.8pt}{\makebox(0,0){$\Diamond$}}}
\put(564,253){\raisebox{-.8pt}{\makebox(0,0){$\Diamond$}}}
\put(900,707){\raisebox{-.8pt}{\makebox(0,0){$\Diamond$}}}
\put(1237,347){\raisebox{-.8pt}{\makebox(0,0){$\Diamond$}}}
\put(1349,820){\raisebox{-.8pt}{\makebox(0,0){$\Diamond$}}}
\put(160.0,123.0){\rule[-0.200pt]{308.111pt}{0.400pt}}
\put(1439.0,123.0){\rule[-0.200pt]{0.400pt}{177.543pt}}
\put(160.0,860.0){\rule[-0.200pt]{308.111pt}{0.400pt}}
\put(160.0,123.0){\rule[-0.200pt]{0.400pt}{177.543pt}}
\end{picture}
\end{center}
\caption{$R(\tau)$ as a function of $\tau$.The numerical values between parentheses in the figure correspond 
to $C_{0\tau}$ (weak coupling case)}
\end{figure}
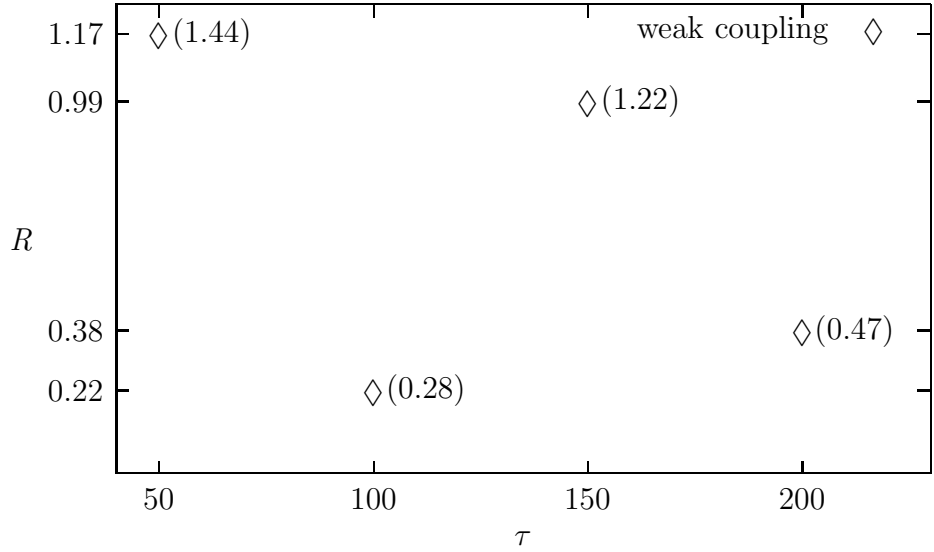

\begin{figure}
\begin{center}
\setlength{\unitlength}{0.240900pt}
\ifx\plotpoint\undefined\newsavebox{\plotpoint}\fi
\sbox{\plotpoint}{\rule[-0.200pt]{0.400pt}{0.400pt}}%
\begin{picture}(1500,900)(0,0)
\sbox{\plotpoint}{\rule[-0.200pt]{0.400pt}{0.400pt}}%
\put(160.0,676.0){\rule[-0.200pt]{4.818pt}{0.400pt}}
\put(140,676){\makebox(0,0)[r]{0.08}}
\put(1419.0,676.0){\rule[-0.200pt]{4.818pt}{0.400pt}}
\put(160.0,492.0){\rule[-0.200pt]{4.818pt}{0.400pt}}
\put(140,492){\makebox(0,0)[r]{0.07}}
\put(1419.0,492.0){\rule[-0.200pt]{4.818pt}{0.400pt}}
\put(1237.0,123.0){\rule[-0.200pt]{0.400pt}{4.818pt}}
\put(1237,82){\makebox(0,0){200}}
\put(1237.0,840.0){\rule[-0.200pt]{0.400pt}{4.818pt}}
\put(900.0,123.0){\rule[-0.200pt]{0.400pt}{4.818pt}}
\put(900,82){\makebox(0,0){150}}
\put(900.0,840.0){\rule[-0.200pt]{0.400pt}{4.818pt}}
\put(564.0,123.0){\rule[-0.200pt]{0.400pt}{4.818pt}}
\put(564,82){\makebox(0,0){100}}
\put(564.0,840.0){\rule[-0.200pt]{0.400pt}{4.818pt}}
\put(227.0,123.0){\rule[-0.200pt]{0.400pt}{4.818pt}}
\put(227,82){\makebox(0,0){50}}
\put(227.0,840.0){\rule[-0.200pt]{0.400pt}{4.818pt}}
\put(160.0,123.0){\rule[-0.200pt]{308.111pt}{0.400pt}}
\put(1439.0,123.0){\rule[-0.200pt]{0.400pt}{177.543pt}}
\put(160.0,860.0){\rule[-0.200pt]{308.111pt}{0.400pt}}
\put(160.0,123.0){\rule[-0.200pt]{0.400pt}{177.543pt}}
\put(19,491){\makebox(0,0){                      {\it R }              }}
\put(799,21){\makebox(0,0){$\tau$}}
\sbox{\plotpoint}{\rule[-0.400pt]{0.800pt}{0.800pt}}%
\sbox{\plotpoint}{\rule[-0.200pt]{0.400pt}{0.400pt}}%
\put(1279,820){\makebox(0,0)[r]{strong coupling}}
\sbox{\plotpoint}{\rule[-0.400pt]{0.800pt}{0.800pt}}%
\put(227,676){\raisebox{-.8pt}{\makebox(0,0){$\Diamond$}}}
\put(564,676){\raisebox{-.8pt}{\makebox(0,0){$\Diamond$}}}
\put(900,676){\raisebox{-.8pt}{\makebox(0,0){$\Diamond$}}}
\put(1237,492){\raisebox{-.8pt}{\makebox(0,0){$\Diamond$}}}
\put(1349,820){\raisebox{-.8pt}{\makebox(0,0){$\Diamond$}}}
\sbox{\plotpoint}{\rule[-0.200pt]{0.400pt}{0.400pt}}%
\put(160.0,123.0){\rule[-0.200pt]{308.111pt}{0.400pt}}
\put(1439.0,123.0){\rule[-0.200pt]{0.400pt}{177.543pt}}
\put(160.0,860.0){\rule[-0.200pt]{308.111pt}{0.400pt}}
\put(160.0,123.0){\rule[-0.200pt]{0.400pt}{177.543pt}}
\end{picture}
\end{center}
\caption{$R(\tau)$ as a function of $\tau$ where $C_{0\tau}=0.79$ (strong coupling case)}
\end{figure}
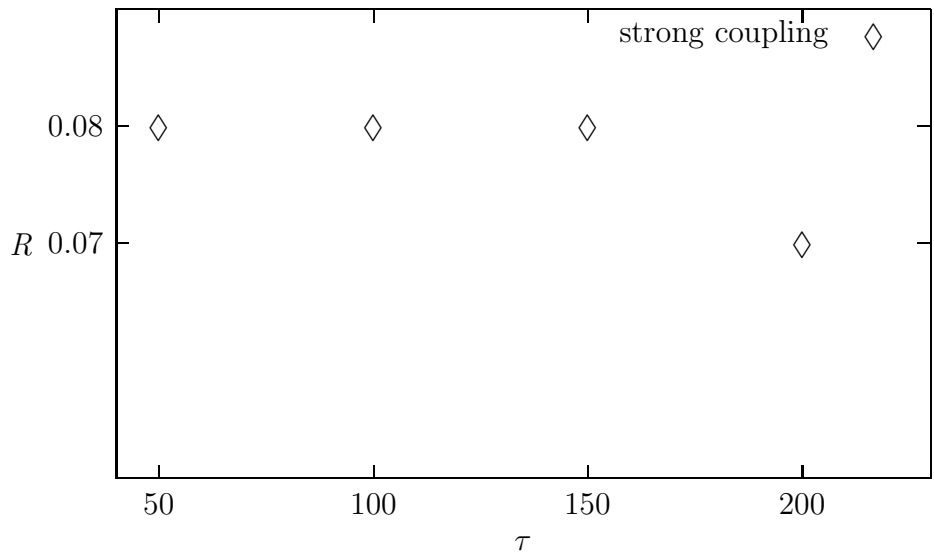


\begin{thebibliography}{99}



\bibitem{tm} L. Mandelstam and I. Tamm, J. Phys. (USSR) {\bf9} (1945) 249  

\bibitem{lv} L. Vaidman, Am. J. Phys. {\bf60} (1992) 182

\bibitem{ju} J. Uffink, Am. J. Phys. {\bf61} (1993) 935 

\bibitem{ml} N. Margolus and L. B. Levitin, Physica D {\bf120} (1998) 188

\bibitem{lt} L. B. Levitin and T. Toffoli, Phys. Rev. Lett.{\bf103} (2009) 160502

\bibitem{jk} P. J. Jones and P. Kok, Phys. Rev. A {\bf82} (2010) 022107

\bibitem{aa} J. Anandan and Y. Aharonov, Phys. Rev. Lett. {\bf65} (1990) 1697

\bibitem{pp} P. Pfeifer, Phys. Rev. Lett. {\bf70} (1993) 3365

\bibitem{and} O. Andersson and H. Heydari, J. Phys. A: Math. Theor. {\bf47} (2014) 215301 

\bibitem{deflu} S. Deffner and E. Lutz, J. Phys. A: Math. Theor. {\bf46} (2013) 335302

\bibitem{deflu1} S. Deffner and E. Lutz, Phys. Rev. Lett. {\bf111} (2013) 010402 

\bibitem{pgi} P. Poggi et al., EPL {\bf104} (2013) 40005

\bibitem{cnv} T. Caneva et al., Phys. Rev. Lett. {\bf103} (2009) 240501  

\bibitem{hgf} G. Hegerfeldt, Phys. Rev. Lett. {\bf111} (2013) 260501  

\bibitem{def} S. Deffner, J. Phys. B: At. Mol. Opt. Phys. {\bf47} (2014) 145502

\bibitem{bas} M. G. Bason et al., Nature Phys. {\bf8} (2012) 147-152  

\bibitem{tad} M. M. Taddei et al.,  Phys. Rev. Lett. {\bf110} (2013) 050402 

\bibitem{cap} A. del Campo et al., Phys. Rev. Lett. {\bf110} (2013) 050403  

\bibitem{bar} E. Barouch, B.M. McCoy and M. Dresden, Phys. Rev. A {\bf2} (1970) 1075

\bibitem{kat} S. Katsura, Phys. Rev. {\bf127} (1962) 1508 

\bibitem{jt} J. Richert and T. Khalil, Physica  B {\bf407} (2012) 729

\end{thebibliography}
\end{document}